\documentclass[12pt]{article}
\usepackage{latexsym,amsmath}
\input amssym
\begin{document}
\renewcommand{\theequation}{\thesection.\arabic{equation}}
\def\prg#1{\medskip{\bf #1}}
\def\lra{\leftrightarrow}        \def\Ra{\Rightarrow}
\def\nin{\noindent}              \def\pd{\partial}
\def\dis{\displaystyle}          \def\dfrac{\dis\frac}
\def\grl{{GR$_\Lambda$}}         \def\vsm{\vspace{-10pt}}
\def\Lra{{\Leftrightarrow}}      \def\ads3{AdS$_3$}
\def\cs{{\scriptscriptstyle \rm CS}}  \def\ads3{{\rm AdS$_3$}}
\def\Leff{\hbox{$\mit\L_{\hspace{.6pt}\rm eff}\,$}}
\def\bull{\raise.25ex\hbox{\vrule height.8ex width.8ex}}
\def\Tr{\hbox{\rm Tr\hspace{1pt}}}
\def\bF{{\bar F}}               \def\bt{{\bar\tau}}
\def\gp{($\Bbb{P}$)}            \def\tn{\tilde\nabla}
\def\tcR{\tilde{\cal R}}        \def\inn{\,\rfloor\,}

\def\D{{\Delta}}      \def\bC{{\bar C}}     \def\bT{{\bar T}}
\def\bH{{\bar H}}     \def\bL{{\bar L}}     \def\bI{{\bar I}}
\def\hO{{\hat O}}     \def\hG{{\hat G}}     \def\tG{{\tilde G}}
\def\cL{{\cal L}}     \def\cM{{\cal M }}    \def\cE{{\cal E}}
\def\cA{{\cal A}}     \def\cI{{\cal I}}     \def\cC{{\cal C}}
\def\cF{{\cal F}}     \def\hcF{\hat{\cF}}   \def\bcF{{\bar\cF}}
\def\cH{{\cal H}}     \def\hcH{\hat{\cH}}   \def\bcH{{\bar\cH}}
\def\cK{{\cal K}}     \def\hcK{\hat{\cK}}   \def\bcK{{\bar\cK}}
\def\cO{{\cal O}}     \def\hcO{\hat{\cal O}} \def\tR{{\tilde R}}
\def\cB{{\cal B}}     \def\cQ{{\cal Q}}

\def\G{\Gamma}        \def\S{\Sigma}        \def\L{{\mit\Lambda}}
\def\a{\alpha}        \def\b{\beta}         \def\g{\gamma}
\def\d{\delta}        \def\m{\mu}           \def\n{\nu}
\def\th{\theta}       \def\k{\kappa}        \def\l{\lambda}
\def\vphi{\varphi}    \def\ve{\varepsilon}  \def\p{\pi}
\def\r{\rho}          \def\Om{\Omega}       \def\om{\omega}
\def\s{\sigma}        \def\t{\tau}          \def\eps{\epsilon}
\def\ups{\upsilon}    \def\tom{{\tilde\om}} \def\bw{{\bar w}}

\def\nab{\nabla}      \def\tnab{{\tilde\nabla}}
\def\Th{\Theta}       \def\cT{{\cal T}}    \def\cS{{\cal S}}
\def\nn{\nonumber}
\def\be{\begin{equation}}             \def\ee{\end{equation}}
\def\ba#1{\begin{array}{#1}}          \def\ea{\end{array}}
\def\bea{\begin{eqnarray} }           \def\eea{\end{eqnarray} }
\def\beann{\begin{eqnarray*} }        \def\eeann{\end{eqnarray*} }
\def\beal{\begin{eqalign}}            \def\eeal{\end{eqalign}}
\def\lab#1{\label{eq:#1}}             \def\eq#1{(\ref{eq:#1})}
\def\bsubeq{\begin{subequations}}     \def\esubeq{\end{subequations}}
\def\bitem{\begin{itemize}}           \def\eitem{\end{itemize}}

\title{Electric field in 3D gravity with torsion}

\author{M. Blagojevi\'c and B. Cvetkovi\'c
\footnote{Email addresses: {\tt
                   mb@phy.bg.ac.yu, cbranislav@phy.bg.ac.yu}} \\
Institute of Physics, P. O. Box 57, 11001 Belgrade, Serbia}
\date{}
\maketitle
\begin{abstract}
It is shown that in static and spherically symmetric configurations
of the system of Maxwell field coupled to 3D gravity with torsion, at
least one of the Maxwell field components has to vanish. Restricting
our attention to the electric sector of the theory, we find an
interesting exact solution, corresponding to the azimuthal electric
field. Its geometric structure is to a large extent influenced by the
values of two different central charges, associated to the asymptotic
AdS structure of spacetime.
\end{abstract}

\section{Introduction}
\setcounter{equation}{0}

Three-dimensional (3D) general relativity (GR) has been used for
nearly three decades as a theoretical laboratory for exploring basic
features of the gravitational dynamics \cite{1}. Among a number of
outstanding results in this field, the discovery of the
Ba\~nados, Teitelboim and Zanelli (BTZ) black
hole \cite{2} was of particular importance, as it resulted in a
significant influence on our understanding of the geometric and
quantum structure of gravity.

In the early 1990s, Mielke and Baekler proposed a new geometric
framework for 3D gravity, in which Riemannian geometry of spacetime was
replaced by the more general, Riemann-Cartan geometry \cite{3}. In this
approach, the gravitational dynamics is characterized by both the
torsion and the curvature \cite{4,5}. Recent developments along these
lines reveal a respectable dynamical content of 3D gravity with
torsion, characterized, in particular, by the existence of the
conformal asymptotic structure with two different central charges, the
BTZ black hole with torsion, the Chern-Simons formulation and the
supersymmetric extension \cite{6,7,8,9}.

The first electrically charged solution in Riemannian 3D gravity was
found in \cite{2}. Later studies of the problem led to a rather
comprehensive analysis of the Einstein-Maxwell dynamics
\cite{10,11,12,13,14}. The purpose of the present paper is to start
similar investigations in 3D gravity with torsion, with a focus on the
electric sector of the theory.

The layout of the paper is as follows. In section 2, we derive the
general field equations of the system of Maxwell field coupled to 3D
gravity with torsion. In section 3, we use these equations to prove a
specific no-go theorem, saying that in static and spherically symmetric
field configurations, at least one component of the Maxwell field has
to vanish. In section 4, we restrict our attention to the electric
sector of the theory and show that, in contrast to Riemannian theory,
dynamically allowed configurations with a radial electric field are
trivial, as the radial field has to be constant. In section 5, we find
an interesting solution generated by the azimuthal electric field, a
generalization of the solution found by Cataldo \cite{14} in Riemannian
GR with a cosmological constant. The solution is shown to have
vanishing conserved charges (energy, angular momentum and electric
charge), while its geometric structure is determined by the central
charges, associated to the asymptotic AdS sector of spacetime. Finally,
appendices contain some technical material.

Our conventions are given by the following rules: the Latin indices
$(i,j,k,...)$ refer to the local Lorentz frame, the Greek indices
$(\m,\n,\l,...)$ refer to the coordinate frame, and both run over
0,1,2; the metric components in the local Lorentz frame are
$\eta_{ij}=(+,-,-)$; totally antisymmetric tensor $\ve^{ijk}$ and the
related tensor density $\ve^{\m\n\r}$ are both normalized so that
$\ve^{012}=1$.

\section{Maxwell field in 3D gravity with torsion} 
\setcounter{equation}{0}

Theory of gravity with torsion can be naturally described as a
Poincar\'e gauge theory (PGT), with an underlying spacetime structure
corresponding to Riemann-Cartan geometry \cite{4,5}.

Basic gravitational variables in PGT are the triad field $b^i$ and
the Lorentz connection $A^{ij}=-A^{ji}$ (1-forms). The corresponding
field strengths are the torsion and the curvature:
$T^i:=db^i+A^i{_m}\wedge b^m$, $R^{ij}:=dA^{ij}+A^i{_m}\wedge A^{mj}$
(2-forms). In 3D, we can simplify the notation by introducing
$A^{ij}=:-\ve^{ij}{_k}\om^k$ and $R^{ij}=:-\ve^{ij}{_k}R^k$, which
yields:
\be
T^i=db^i+\ve^i{}_{jk}\om^j\wedge b^k \, ,\qquad
R^i=d\om^i+\frac{1}{2}\,\ve^i{}_{jk}\om^j\wedge\om^k\, .   \lab{2.1}
\ee

The covariant derivative $\nabla(\om)$ acts on a general
tangent-frame spinor/tensor in accordance with its
spinorial/tensorial structure; when $X$ is a form, $\nabla
X:=\nabla\wedge X$.

PGT is characterized by a useful identity:
\bsubeq\lab{2.2}
\be
\om^i\equiv\tom^i+K^i\, ,                                  \lab{2.2a}
\ee
where $\tom^i$ is the Levi-Civita (Riemannian) connection, and $K^i$
is the contortion 1-form, defined implicitly by
\be
T^i=:\ve^i{}_{mn}K^m\wedge b^n\, .                         \lab{2.2b}
\ee
Using this identity, one can express the curvature $R_i=R_i(\om)$ in
terms of its {\it Riemannian\/} piece $\tR_i=R_i(\tom)$ and the
contortion $K_i$:
\be
2R_i\equiv 2\tR_i+2\tnab K_i+\ve_{imn}K^m\wedge K^n\, .    \lab{2.2c}
\ee
\esubeq

The antisymmetry of the Lorentz connection $A^{ij}$ implies that the
geometric structure of PGT corresponds to Riemann-Cartan geometry, in
which $b^i$ is an orthonormal coframe, $g:=\eta_{ij}b^i\otimes b^j$ is
the metric of spacetime, and $\om^i$ is the Cartan connection.

In local coordinates $x^\m$, we can write $b^i=b^i{_\m}dx^\m$, the
frame dual to $b^i$ reads $h_i=h_i{^\m}\pd_\m$ and satisfies the
property $h_i \inn b^j=h_i{^\m}b^j{_\m}=\d^j_i$, where $\inn$ is the
interior product. In what follows, we will omit the wedge product sign
$\wedge$ for simplicity.

\subsection*{Lagrangian and the field equations}

General gravitational dynamics in Riemann-Cartan spacetime is
determined by Lagrangians which are at most quadratic in field
strengths. Omitting the quadratic terms, we arrive at the {\it
topological\/} Mielke-Baekler (MB) model for 3D gravity \cite{3}:
\bsubeq\lab{2.3}
\be
L_0=2ab^i R_i-\frac{\L}{3}\,\ve_{ijk}b^i b^j b^k\
    +\a_3L_\cs(\om)+\a_4 b^i T_i\, .                       \lab{2.3a}
\ee
Here, $a=1/16\pi G$ and $L_\cs(\om)$ is the Chern-Simons Lagrangian for
the Lorentz connection, $L_\cs(\om)=\om^id\om_i
+\frac{1}{3}\ve_{ijk}\om^i\om^j\om^k$. The MB model is a natural
generalization of GR with a cosmological constant (\grl).

The complete dynamics includes also the contribution of matter fields,
minimally coupled to gravity. We focus our attention to the case when
matter is represented by the Maxwell field:
\be
L=L_0+L_M\, ,\qquad L_M:=-\frac{1}{4}F{}^*F\, ,                                              \lab{2.3b}
\ee
\esubeq
where $F=dA$.

By varying $L$ with respect to $b^i$ and $\om^i$, one obtains the
gravitational field equations:
\bea
&&2aR_i+2\a_4T_i-\L\ve_{ijk}b^jb^k=\Th_i\, ,               \nn\\
&&2\a_3R_i+2aT_i+\a_4\ve_{ijk}b^jb^k=\S_i\, ,
\eea
where $\Th_i:=-\d L_M/\d b^i$ and $\S_i:=-\d L_M/\d\om^i$ are the
energy-momentum and spin currents (2-forms) of matter. The Maxwell
field currents are given by
$$
\Th_i=\frac{1}{2}\left[F(h_i\inn{}^*F)-(h_i\inn F){}^*F\right]\, ,
\qquad \S_i=0\, .
$$
In the nondegenerate sector with $\D:=\a_3\a_4-a^2\neq 0$, these
equations can be rewritten as \bsubeq\lab{2.5} \bea
&&2T_i-p\ve_{ijk}b^jb^k=u\Th_i\, ,                         \lab{2.5a}\\
&&2R_i-q\ve_{ijk}b^jb^k=-v\Th_i\, ,                        \lab{2.5b}
\eea
\esubeq
where
\bea
&&p:=\frac{\a_3\L+\a_4 a}{\D}\, ,\qquad
  q:=-\frac{(\a_4)^2+a\L}{\D}\, ,                          \nn\\
&&u:=\frac{\a_3}{\D}\, ,\qquad  v:=\frac{a}{\D}\, .        \nn
\eea

Introducing the energy-momentum tensor of matter by
$\cT^k{_i}:={}^*(b^k\Th_i)$, the matter current $\Th_i$ can be
expressed as follows:
\bsubeq\lab{2.6}
\bea
&&\Th_i=\frac{1}{2}\left(\cT^k{_i}\ve_{kmn}\right)b^mb^n
       =\ve_{imn}t^mb^n\, ,                                  \\
&&t^m:=-\left(\cT^m{_k}-\frac{1}{2}\d^m_k\cT\right)b^k\, ,
\eea
where $\cT=\cT^k{_k}$. The Maxwell energy-momentum tensor reads:
\be
\cT^k{_i}=-F^{km}F_{im}+\frac{1}{4}\d^k_iF^2\, ,
\ee
\esubeq
with $F^2=F^{mn}F_{mn}$,

These results can be used to simplify the field equations \eq{2.5}.
Indeed, if we substitute the above $\Th_i$ into \eq{2.5a} and compare
the result with \eq{2.2b}, we find the following form of the
contortion:
\bsubeq\lab{2.7}
\be
K^m=\frac{1}{2}(pb^m+ut^m)\, .                             \lab{2.7a}
\ee
After that, we can rewrite the second field equation \eq{2.5b} as
\be
2R_i=q\ve_{imn}b^mb^n-v\ve_{imn}t^m b^n\, ,                \lab{2.7b}
\ee
where the Cartan curvature $R_i$ is calculated using the identity
\eq{2.2c}:
\be
2R_i=2\tR_i+u\tnab t_i
 +\ve_{imn}\left(\frac{p^2}{4}b^mb^n
 +\frac{up}2\,t^mb^n+\frac{u^2}4t^mt^n\right)\, .          \lab{2.7c}
\ee \esubeq In this form of the gravitational field equations, the
role of the Maxwell field as a source of gravity is clearly
described by the 1-form $t^i$.

The Maxwell field equations take the standard form:
\be
d{}^*F=0\, .                                               \lab{2.8}
\ee
These equations, together with a suitable set of boundary conditions,
define the complete dynamics of both the gravitational and the
Maxwell field.

\section{A no-go theorem} 
\setcounter{equation}{0}

In order to explore basic dynamical features of the system of
Maxwell field coupled to 3D gravity with torsion, we begin by
looking at \emph{static} and \emph{spherically symmetric} field
configurations. Using the Schwarz\-schild-like coordinates
$x^\m=(t,r,\vphi)$, we make the following ansatz for the triad
field,
\be
b^0=Ndt\, ,\qquad b^1=B^{-1}dr\, ,\qquad b^2=Kd\vphi\,,\lab{3.1}
 \ee
 and for the Maxwell field:
 \be
F=E_rb^0b^1-Hb^1b^2+E_\vphi b^2b^0\, . \lab{3.2}
\ee
Here, $N,B,K$
and $E_r,H,E_\vphi$ are the unknown functions of the radial coordinate
$r$.

The Maxwell equations \eq{2.8} read:
\bsubeq
\be
E_r'B+\g E_r=0\, ,\qquad H'B+\a H=0\, ,\qquad
E_\vphi=E_\vphi(r)\, ,                                     \lab{3.3a}
\ee
where $\a,\g$ are components of the Riemannian connection, defined in
Appendix A, and $E_\vphi$ remains an arbitrary function of $r$. The
corresponding first integrals are
\be
E_rK=Q_1\,,\qquad HN=Q_3\, ,                               \lab{3.3b}
\ee
\esubeq
where $Q_1$ and $Q_3$ are constants.

Next, we calculate the energy-momentum tensor
\be
\cT^i{_j}=\frac{1}{2}
  \left( \ba{ccc}
         \dis E_r^2+E_\vphi^2+H^2 & 2E_\vphi H & 2E_rH \\
         -2E_\vphi H & E_r^2-E_\vphi^2-H^2 & -2E_rE_\vphi \\
         -2E_rH& -2E_rE_\vphi &-E_r^2+E_\vphi^2-H^2
         \ea\right)\, ,
\ee
and find the expression for $t^i$:
\bea
&&t^0=-\left(\frac{\cT}{2}+H^2\right)b^0
      -E_\vphi Hb^1-E_rHb^2\, ,                            \nn\\
&&t^1=E_\vphi Hb^0-\left(\frac{\cT}{2}-E_\vphi^2\right)b^1
      +E_rE_\vphi b^2\, ,                                  \nn\\
&&t^2=E_rHb^0+E_rE_\vphi b^1
      -\left(\frac{\cT}{2}-E_r^2\right)b^2\,,
\eea
where $\cT=\cT^k{}_k=\dis\frac{1}{2}\left(E_r^2+E_\vphi^2-H^2\right)$.

Technical details leading to the explicit form of the
gravitational field equations \eq{2.7b} are summarized in appendix
A. All the components of these equations can be conveniently
divided in two sets: those with $(i,m,n)= (0,1,2),(2,0,1),(1,2,0)$
are called diagonal, all the others are nondiagonal. Introducing
$$
V:=v+up/2
$$
to simplify the notation, the nondiagonal equations read:

\bsubeq\lab{3.6} \bea &&VE_rH-\frac{u^2}{4}E_rH\cT
  =u\left(-\frac{1}{2}\g E_r^2+\frac{1}{2}E_\vphi E_\vphi'B
   +\a E_\vphi^2-\frac{1}{2}\a H^2\right)\, ,              \lab{3.6a}\\
&&VE_\vphi H-\frac{u^2}{4}E_\vphi H\cT
  =-uE_r E_\vphi\a\, ,                                     \lab{3.6b}\\
&&VE_rE_\vphi-\frac{u^2}{4}E_rE_\vphi\cT
  =uH(E_\vphi'B+\a E_\vphi)\, ,                            \lab{3.6c}\\
&&VE_\vphi H-\frac{u^2}{4}E_\vphi H\cT
  =uE_r(E_\vphi'B+\g E_\vphi)\, ,                          \lab{3.6d}\\
&&VE_rH-\frac{u^2}{4}E_r H\cT
  =-u\left(\frac{1}{2}\g E_r^2+\frac{1}{2}E_\vphi E_\vphi'B
   +\g E_\vphi^2+\frac12\a H^2\right)\, ,                  \lab{3.6e}\\
&&VE_rE_\vphi-\frac{u^2}{4}E_rE_\vphi\cT
  =-uE_\vphi H\g\, ,                                       \lab{3.6f}
\eea
\esubeq
while the diagonal ones are:
\bsubeq\lab{3.7}
\bea
&&-2(\g'B+\g^2)=2\Leff-V(\cT+H^2)+\frac{u^2}{4}\cT\left(\frac{3}{2}\cT
  +H^2\right)-uE_rH\a\, ,                                  \lab{3.7a}\\
&&-2\a\g=2\Leff-V(\cT-E_\vphi^2)
  +\frac{u^2}4\cT\left(\frac{3}{2}\cT-E_\vphi^2\right)
  -uE_rH(\g-\a)\, ,                                        \lab{3.7b}\\
&&-2(\a'B+\a^2)=2\Leff-V(\cT-E_r^2)
  +\frac{u^2}{4}\cT\left(\frac{3}{2}\cT-E_r^2\right)
  +uE_rH\g\, ,                                             \lab{3.7c}
\eea
\esubeq
where $\Leff:=q-p^2/4$. These equations are invariant under the
\emph{duality mapping}
\bea
&&\a\to\g\, ,\qquad \g\to\a\, ,                            \nn\\
&&E_r\to iH\, ,\qquad H\to iE_r\, ,                        \lab{3.8}
\eea
which defines a useful correspondence between different solutions.
The duality mapping has the same form as in \grl\ \cite{14}.

In the case $u=0$, one immediately concludes that
$$
E_rE_\vphi=0\,,\qquad  E_rH=0\,,\qquad E_\vphi H=0\, .
$$
This is a specific no-go theorem, which holds in \grl\ \cite{14};
it states that configurations with two nonvanishing components of
the Maxwell field are dynamically not allowed.

Let us now return to the general case with $u\neq 0$. Analyzing
the above gravitational field equations, one obtains the general
no-go theorem (appendix B):
\be
E_rE_\vphi H=0\, .\lab{3.9}
\ee
 \bitem
  \item[\bull] In any static and spherically
symmetric configuration, it is dynamically impossible to have
three nonvanishing components of the Maxwell field.
\eitem

The theorem implies that at least one component of the Maxwell field
has to vanish. Motivated by this result, we now turn our attention to
exploring static and spherically symmetric solutions associated with an
\emph{electric} Maxwell field, specified by $H=0$.

\section{Dynamics in the electric sector} 
\setcounter{equation}{0}

The electric sector of the Maxwell field is defined by
$$
F=E_rb^0b^1+E_\vphi b^2b^0\, .
$$
The field equations are significantly simplified. The nondiagonal
equations take the form
\bsubeq\lab{4.1}
\bea
&&0=-u\left(-\frac{1}{2}\g E_r^2+\frac{1}{2}E_\vphi E_\vphi'B
    +\a E_\vphi^2\right),                                  \lab{4.1a}\\
&&0=uE_r E_\vphi\a\, ,                                     \lab{4.1b}\\
&&VE_rE_\vphi
  -\frac{u^2}{4}E_rE_\vphi\cT=0\, ,                        \lab{4.1c}\\
&&0=-uE_r(E_\vphi'B+\g E_\vphi)\, ,                        \lab{4.1d}\\
&&0=u\left(\frac{1}{2}\g E_r^2+\frac{1}{2}E_\vphi E_\vphi'B
    +\g E_\vphi^2\right)\, ,                               \lab{4.1e}
\eea
\esubeq
while the diagonal ones are:
\bsubeq\lab{4.2}
\bea
&&-2(\g'B+\g^2)=2\Leff-V\cT
  +\frac{3u^2}{8}\cT^2\, ,                                 \lab{4.2a}\\
&&-2\a\g=2\Leff-V(\cT-E_\vphi^2)
  +\frac{u^2}4\cT\left(\frac{3}{2}\cT-E_\vphi^2\right)\, , \lab{4.2b}\\
&&-2(\a'B+\a^2)=2\Leff-V(\cT-E_r^2)
  +\frac{u^2}{4}\cT\left(\frac{3}{2}\cT-E_r^2\right)\, .   \lab{4.2c}
\eea
\esubeq

The analysis of these equations leads to the conclusion that the only
interesting configuration is the one defined by the azimuthal electric
field $E_\vphi$.

\subsection*{\boldmath $E_rE_\vphi\ne 0$}

In this case, equations \eq{4.1b} and \eq{4.1c} imply
$$
\a=0\, ,\qquad V-\frac{u^2}{4}\cT=0\, ,
$$
while \eq{4.2c} yields
$$
\Leff+\left(\frac{V}{u}\right)^2=0\, .
$$
Using $V=v+up/2$ and the identity $ap+\a_3q+\a_4=0$, the last relation
leads to $\a_3\a_4-a^2=0$, which is in contradiction with the property
$\D\ne 0$, adopted in section 2. Hence, our assumption $E_rE_\vphi\ne
0$ cannot be true, i.e. at least one of the components $E_r,E_\vphi$
must vanish.

\subsection*{\boldmath $E_r\ne 0, E_\vphi=0$}

In this case, there exists only the radial electric field. Equation
\eq{4.1a} yields $\g=0$, which implies
$$
K= {\rm const.}\quad \Ra \quad E_r = {\rm const.}
$$
Equation \eq{4.2a} determines the value of $\cT=E_r^2/2$. Combining
\eq{4.2a} and \eq{4.2c}, we obtain
$$
\a'+\a^2=-2\Leff-\frac{u^2}8\cT^2=:\k^2\, ,                                          \lab{4.8}
$$
where the radial coordinate is chosen so that $B=1$, for simplicity.
The solution of this equation depends on the sign of $\k^2$.
\bitem\vsm
\item[a)] For $\k^2>0$, we have $N=C\cosh\k(r+r_0)$, where $r_0$ and
$C$ are integration constants. \vsm
\item[b)] For $\k^2=0$, we find $N=C(r+r_0)$. \vsm
\item[c)] For $\k^2<0$ we find $N=C\sin|\k|(r+r_0)$.
\eitem\vsm

All of these solutions are characterized by a constant electric field,
hence, they are of no physical interest. As we shall see in the next
section, the last case, defined by a sole azimuthal electric field,
leads to a nontrivial dynamical situation.

\section{Solution with azimuthal electric field} 
\setcounter{equation}{0}

Since the Maxwell field equations do not impose any restriction on
the azimuthal electric field, it is completely determined by the
gravitational field equations.

The non-diagonal gravitational field equations are very simple:
\bsubeq
\be
E'_\vphi B+2\a E_\vphi=0\, ,\qquad E'_\vphi B+2\g E_\vphi=0\, .
\ee
They imply
\bea
&&\a=\g\quad\Ra\quad N=C_1K\, ,                            \nn\\
&&E_\vphi K^2=Q_2\, ,
\eea
\esubeq
where $C_1$ and $Q_2$ are the integration constants.

In the set of the diagonal field equations \eq{4.2}, the first two
take the form
\bsubeq\lab{5.2}
\bea
&&-2(\g'B+\g^2)=2\Leff-V\cT+\frac{3u^2}{8}\cT^2\, ,        \lab{5.2a}\\
&&-2\a\g=2\Leff+V\cT-\frac{u^2}{8}\cT^2\, ,                \lab{5.2b}
\eea
\esubeq
while the third one is equivalent to \eq{5.2a}, since $\a=\g$.
It is convenient to fix the radial coordinate by choosing
\be
K=r\quad\Ra\quad E_\vphi=\frac{Q_2}{r^2} \, ,              \lab{5.3}
\ee
whereupon the field equations take the form
\bsubeq\lab{5.4}
\bea
&&B^2=-\Leff r^2-V\frac{Q_2^2}{4r^2}
      +\frac{u^2}{64}\frac{Q_2^4}{r^6}\, ,                 \lab{5.4a}\\
&&BB'=-r\Leff+V\frac{Q_2^2}{4r^3}
      -\frac{3u^2}{64}\frac{Q_2^4}{r^7}\, .                \lab{5.4b}
\eea
\esubeq
These two equations are consistent with each other, and they
determine $B^2$.

The above expressions for $N,B,K$ and $E_\vphi$ represent a complete
solution describing the azimuthal electric field in 3D gravity with
torsion, for any value of $\Leff$. In what follows, we shall restrict
our considerations to the case of negative $\Leff$,
$$
\Leff=-\frac{1}{\ell^2}\, ,
$$
which corresponds asymptotically to an AdS configuration, suitable for
the calculation of the conserved charges. Using $C_1=1/\ell$, the
solution can be written in the form
\bsubeq\lab{5.5}
\bea
&&b^0=\frac{r}{\ell}dt\, ,\qquad b^1=B^{-1}dr\, ,
  \qquad b^2=rd\vphi\, ,                                   \nn\\
&&\tom^0=-Bd\vphi\, ,\qquad \tom^1=0\, ,
  \qquad \tom^2=-\frac{B}{\ell} dt\, ,                     \nn\\
&&t^0=-\frac{Q_2^2}{4\ell r^3}dt\, ,
  \qquad t^1=\frac{3Q_2^2B}{4r^4}dr\,,
  \qquad t^2=-\frac{Q_2^2}{4r^3}d\vphi\, ,                 \nn\\
&&\om^i=\tom^i+\frac{1}{2}\left(pb^i+ut^i\right)\, ,
\eea
and
\be
F=\frac{Q_2}{\ell} d\vphi dt\,.
\ee
\esubeq
For $Q_2=0$, the solution reduces to the black hole vacuum.

\subsection*{Conserved charges}

A deeper insight into the nature of the exact solution \eq{5.5} is
achieved by calculating the values of its conserved charges.

We start by choosing the \emph{asymptotic conditions} at spatial
infinity so that (i) the fields  $b^i$, $\om^i$ and $F$ belong to the
family \eq{5.5}, parametrized by $Q_2$, and (ii) the corresponding
asymptotic symmetries have well-defined canonical generators.

According to (i), the asymptotic form of the fields reads:

\bea
&&b^i{_\m}\sim\left( \ba{ccc}
  \dis\frac{r}{\ell} & 0 & 0 \\
  0 & \dis\frac{\ell}{r}+V\frac{\ell^3Q_2^2}{8r^5} & 0 \\
  0 & 0 & r
                      \ea
             \right)\, ,                                   \nn\\
&&\om^i{_\m}\sim \left( \ba{ccc}
  \dis\frac{pr}{2\ell}-\frac{uQ_2^2}{8\ell r^3} &  0 &
     -\dis\frac{r}{\ell}+V\frac{\ell Q_2^2}{8r^3} \\
  0 & \dis\frac{p\ell}{2r}+\frac{3uQ_2^2}{8\ell r^3} & 0 \\
 -\dis\frac{r}{\ell^2}+V\frac{Q_2^2}{8r^3} & 0 &
      \dis\frac{pr}{2}-\frac{uQ_2^2}{8\ell r^3}
                        \ea
                 \right) \, ,                              \lab{5.6}\\
&&F\sim \frac{Q_2}\ell d\vphi dt\, .                       \nn
\eea

Gauge symmetries of the theory are local translations, local Lorentz
rotations and local $U(1)$ transformations, parametrized by $\xi^\m$,
$\th^i$ and $\l$, respectively. The subset of gauge transformations
that respects the adopted asymptotic conditions is defined by the
\emph{asymptotic parameters}, which have the following form:
\be
\xi^\m=(\ell T_0,0,S_0)\, , \qquad
\th^i=(0,0,0)\, ,\qquad  \l=\l(t,r,\vphi)\, .              \lab{5.7}
\ee
Here $T_0$ and $S_0$  are constant parameters associated with the rigid
time translations and axial rotations, respectively, while the $U(1)$
parameter $\l$ remains local, since the asymptotic conditions
\eq{5.6} do not restrict its form.

We are now ready to calculate  the asymptotic charges by using the
standard canonical approach. Technical details leading to the
construction of the canonical generator $G$ are summarized in appendix
C. Since $G$ acts on basic dynamical variables via the Poisson
brackets, it must be differentiable. If this is not the case, the form
of $G$ can be improved by adding a suitable surface term. However, it
turns out that under the adopted asymptotic conditions \eq{5.6},  $G$
is differentiable automatically, without adding any surface term.
Consequently, the canonical charges corresponding to the asymptotic
parameters \eq{5.7}, the energy $E$, the angular momentum $M$ and the
electric charge $Q$ of the solution \eq{5.5}, vanish:
\be
E=0\, ,\qquad M=0\, ,\qquad Q=0.                           \lab{5.8}
\ee
The vanishing of the electric charge is an expected result, since the
azimuthal electric does not produce the radial flux. The vanishing of
$E$ and $M$ can be understood by noting that the subleading terms  in
the asymptotic formulas \eq{5.6} are ''too small'' to produce any
nontrivial contribution to the corresponding surface integrals (a
nontrivial contribution would be produced by the subleading terms of
order $r^{-1}$).

\subsection*{Geometric structure}

The metric of the solution \eq{5.5} reads:
\be
ds^2=\frac{r^2}{\ell^2}dt^2-\frac{dr^2}{\dis\frac{r^2}{\ell^2}
     -V\frac{Q_2^2}{4r^2}
     +\frac{u^2}{64}\frac{Q_2^4}{r^6}}-r^2d\vphi^2\, .     \nn
\ee
In the limit $u\to 0$, the metric reduces to the form found by Cataldo
in \grl\ \cite{15}, while for $Q_2=0$, it coincides with the black hole
vacuum.

The scalar Cartan curvature is singular at $r=0$:
$$
R=-\ve^{imn}R_{imn}=-6q+\frac{Q_2^2}{2r^4}\, .
$$

Let us now show that the above solution is regular in the region $r>0$.
We begin by writing $B^2(r)=f(r^4)/r^6$, where
$$
f(x):=\frac{x^2}{\ell^2}-V\frac{Q_2^2}{4}x+\frac{u^2Q_2^4}{64}\, .
$$
For large $r$, the function $B^2$ is positive. The coordinate
singularities are related to the existence of zeros of $f(x)$ for some
real and positive $x$. Equation $f(x)=0$ has real and positive roots if
the following two conditions are fulfilled:
\be
V^2-\frac{u^2}{\ell^2}\ge 0\, ,\qquad V>0\, .              \nn
\ee
Using the relations
$$
V^2-\frac{u^2}{\ell^2}=-\frac{u^2}{\a_3^2}\D\, ,\qquad
V\mp\frac{u}{\ell}=\frac{1}{24\pi\ell\D}c^\mp\, ,
$$
where $c^\mp$ are classical central charges \cite{7}, and recalling
that $\D\ne 0$, the above conditions imply
\bea
&&\D<0\quad\Lra\quad c^-c^+> 0\, ,                        \nn\\
&&\frac{1}{\D}(c^-+c^+)> 0\quad\Ra\quad c^-+c^+<0\, .     \nn
\eea
By combining these relations, one concludes that both central charges
have to be negative. However, the positivity of the BTZ black hole
energy, obtained in the supersymmetric 3D gravity with torsion [6],
implies that both $c^-$ and $c^+$ are positive. Thus, $f(x)$ has no
real and positive zeros, and consequently, $B^2(r)>0$ for all $r>0$.

\bitem
\item[\bull] For $Q_2\ne 0$, the solution \eq{5.5} is regular in
the region $r>0$ and does not have the black hole structure.
\eitem

Although the central charges $c^\mp$ are originally defined by the
asymptotic structure of the AdS sector of spacetime, here, in the
presence of the azimuthal electric field, they have a new
dynamical role, embedded in the geometric properties of the
solution \eq{5.5}. The same phenomenon is observed also in the
solution with self-dual Maxwell field \cite{16}.

\section{Concluding remarks} 

In this paper, we studied exact solutions of the system of electric
field coupled to 3D gravity with torsion.

(1) First, we proved the following theorem: in static and spherically
symmetric configurations, the Maxwell field cannot have all three
components different from zero.

(2) Motivated by this result, we restricted our consideration to
the electric sector of the theory and found an exact solution with
azimuthal electric field. This is the only nontrivial solution in
this sector, and it represents a generalization of the Cataldo
solution, found earlier in Riemannian \grl. The values of its
energy, angular momentum and electric charge all vanish, and its
geometric structure is strongly influenced by the classical
central charges, associated with the asymptotic AdS structure of
spacetime.

\appendix
\section*{Acknowledgements} 

This work was supported by the Serbian Science Foundation.

\section{The field equations} 
\setcounter{equation}{0}

In this appendix, we present some technical details regarding the
structure of the field equations \eq{2.7b}.

In Schwarzschild-like coordinate $(t,r,\vphi)$, the form of the
static and spherically symmetric triad field is defined in \eq{3.1}.
After calculating the Levi-Civita connection,
\bsubeq
\be
\tom^0=-\g b^2\, ,\qquad\tom^1=0\, ,\qquad\tom^2=-\a b^0\,,\nn
\ee
where $\a=BN'/N$, $\g=BK'/K$, we find the Riemannian curvature:
\bea
&&\tR_0=-(\g'B+\g^2)b^1b^2\, ,\qquad\tR_1=-\a\g b^2b^0\, , \nn\\
&&\tR_2=-(\a'B+\a^2)b^0b^1\, .                             \nn
\eea
\esubeq

In the next step, we calculate the expressions $B_i=\ve_{imn}t^mb^n$,
$C_i=\ve_{imn}t^mt^n$ and $\tn t_i$:
\bsubeq
\bea
&&B_0=-E_r Hb^0b^1+\left(\cT+H^2\right)b^1b^2-E_\vphi Hb^2b^0\, ,\nn\\
&&B_1=-E_rE_\vphi b^0b^1+E_\vphi Hb^1b^2
      +\left(\cT-E_\vphi^2\right)b^2b^0\, ,                \nn\\
&&B_2=\left(\cT-E_r^2\right)b^0b^1+E_rHb^1b^2
      -E_rE_\vphi b^2b^0\, ,                               \nn
\eea
\bea
&&C_0=E_rH\cT b^0b^1-\cT\left(\frac{3}{2}\cT+H^2\right)b^1b^2
      +E_\vphi H\cT b^2b^0\, ,                             \nn\\
&&C_1=E_rE_\vphi\cT b^0b^1-E_\vphi H\cT b^1b^2
      -\cT\left(\frac{3}{2}\cT-E_\vphi^2\right)b^2b^0\, ,  \nn\\
&&C_2=-\cT\left(\frac{3}{2}\cT-E_r^2\right)b^0b^1-E_rH\cT b^1b^2
      +E_rE_\vphi\cT b^2b^0\, ,                            \nn
\eea
\bea
&&\tn t_0=\left(-\frac{1}{2}\g E_r^2+\frac{1}{2}E_\vphi E_\vphi'B
          +\a E_\vphi^2-\frac{1}{2}\a H^2\right)b^0b^1
          +E_rH\a b^1b^2-E_r E_\vphi\a b^2b^0\, ,          \nn\\
&&\tn t_1=H(E_\vphi'B+\a E_\vphi)b^0b^1-E_r(E_\vphi'B
          +\g E_\vphi)b^1b^2+E_rH(\g-\a)b^2b^0\, ,         \nn\\
&&\tn t_2=-E_rH\g b^0b^1+\left(\frac{1}{2}\g E_r^2
  +\frac{1}{2}E_\vphi E_\vphi'B+\g E_\vphi^2
  +\frac{1}{2}\a H^2\right)b^1b^2-E_\vphi H\g b^2b^0 \, .  \nn
\eea
\esubeq

The above results completely determine the Cartan curvature
\eq{2.7c}, and lead to the second gravitational field equation in the
form \eq{3.6} and \eq{3.7}.

\section{The proof of {\boldmath $E_rE_\vphi H =0$}} 
\setcounter{equation}{0}

In this appendix, we prove the general no-go theorem formulated at
the end of section 3.

We begin by assuming $E_rE_\vphi H\ne 0$. Then, the consistency of
the first three and the last three nondiagonal equations in \eq{3.6}
is ensured by a single relation:
\be\
E_\vphi'B+ E_\vphi(\a+\g)=0\qquad\Rightarrow \qquad
E_\vphi NK=C_1\, ,                                         \lab{B1}
\ee
where $C_1$ is a constant. With \eq{B1}, the set of equations
\eq{3.6} reduces to:
\bea
&&VE_rH-\frac{u^2}{4}E_rH\cT
  =u\left(-\frac{1}{2}\g E_r^2+\frac{1}{2}\a E_\vphi^2
   -\frac{1}{2}\g E_\vphi^2-\frac{1}{2}\a H^2\right)\, ,   \nn\\
&&VE_\vphi H-\frac{u^2}{4}E_\vphi H\cT
  =-uE_r E_\vphi\a\, ,                                     \nn\\
&&VE_rE_\vphi-\frac{u^2}{4}E_rE_\vphi\cT
  =-uE_\vphi H\g\, ,                                       \lab{B2}
\eea
The consistency of these equations implies:
\be
\a E_r^2=\g H^2\, , \qquad (\g-\a)\cT=0\, ,
\ee
which yields either $\a=\g$ or $\cT=0$. Now, we use this result to
explore the diagonal equations \eq{3.7}.

(a) If $\a=\g$, then  we have $E_r=\eps H$, $\eps=\pm 1$,
equations \eq{B2} and \eq{3.7b} yield
\be
V-\frac{u^2}4\cT=-\eps u\a\, ,\qquad
\Leff+\left(\eps\a-\frac{u\cT}4\right)^2=0\, ,             \nn
\ee
and consequently,
\be
\Leff+\left(\frac{V}{u}\right)^2=0\, .                     \lab{B4}
\ee
This relation leads to $\a_3\a_4-a^2=0$, in contradiction with the
assumption $\D\ne 0$, adopted in section 2. Thus, $\a=\g$ is not
allowed.

(b) If $\cT=0$, equations \eq{B2} imply
\be
VH=-uE_r\a\, ,\qquad  VE_r=-uH\g\, ,                       \nn
\ee
leading to $V^2=u^2\a\g$. Then, \eq{3.7b} reduces to $\a\g=-\Leff$,
and we obtain again \eq{B4}, so that $\cT=0$ is also not allowed.

This completes the proof that the configuration $E_rE_\vphi H\ne 0$ can
not be realized dynamically, which is exactly the content of the no-go
theorem.

\section{Canonical generator} 
\setcounter{equation}{0}

In this appendix, we  construct the canonical generator of gauge
transformations.

Starting with the Lagrangian variables $(b^i{}_\m,\om^i{}_\m,A_\m )$
and the related canonical momenta $(\pi^i{}_\m,\Pi^i{}_\m,\pi^\m)$,
we find the following primary constraints:
\bea
&&\phi_i{}^0:=\pi_i{}^0\approx 0\, , \qquad
\phi_i{}^\a:=\pi_i{}^\a-\a_4\ve^{0\a\b}b_{i\b}\approx 0\, ,\nn\\
&&\Phi_i{}^0:=\Pi_i{}^0\approx 0\, , \qquad
    \Phi_i{}^\a:=\Pi_i{}^\a-\ve^{0\a\b}\left(2a b_{i\b}
                 +\a_3\om_{i\b}\right)\approx 0\, ,        \nn\\
&&\phi:=\pi^0\approx 0\, ,                                 \nn
\eea
The canonical Hamiltonian is linear in unphysical variables, as
expected:
\bea
&&\cH_c= b^i{}_0\cH_i+\om^i{}_0\cK_i
        -A_0\pd_\a\pi^\a+\pd_\a D^\a\, ,                   \nn\\
&&\cH_i=-\ve^{0\a\b}\Bigl(aR_{i\a\b}+\a_4T_{i\a\b}
  -\L\ve_{ijk}b^j{}_\a b^k{}_\b-\frac{1}{2}\Th_{i\a\b}\Bigr)\,,\nn\\
&&\cK_i=-\ve^{0\a\b}\left(aT_{i\a\b}+\a_3R_{i\a\b}
  +\a_4\ve_{ijk}b^j{}_\a b^k{}_\b\right)\, ,               \nn\\
&&D^\a=\ve^{0\a\b}\left[ \om^i{}_0\left( 2ab_{i\b}
   +\a_3\om_{i\b}\right)+\a_4b^i{}_0 b_{i\b}\right]+A_0\pi^\a\,.\nn
\eea
Going over to the total Hamiltonian,
\be
\cH_T=\cH_c+u^i{}_\m\phi_i{}^\m+v^i{}_\m\Phi_i{}^\m+w\pi^0\,,\nn
\ee
we find that the consistency conditions of the  primary constraints
$\pi_i{}^0$, $\Pi_i{}^0$ and $\pi^{0}$ yield the secondary
constraints:
\be
\cH_i\approx0\, , \qquad \cK_i\approx0\, , \qquad
\pd_\a\pi^\a\approx 0\, .                                  \nn
\ee

The consistency of the additional primary constraints $\phi_i{}^\a$
and $\Phi_i{}^\a$, leads to the determination of the multipliers
$u^i{}_\a$ and $v^i{}_\a$, resulting in the final form of the total
Hamiltonian:
\bea
\cH_T&=&\hcH_T+\pd_\a\hat{D}^\a\, ,                        \nn\\
\hcH_T&=&b^i{}_0\hcH_i+\om^i{}_0\hat\cK_i-A_0\pd_\a\pi^\a
+u^i{}_0\pi_i{}^0+v^i{}_0\Pi_i{}^0+w\pi^0\, ,              \nn
\eea
where
 \bea
 &&\hcH_i=\cH_i-\nabla_\b\phi_i{}^\b
  +\ve_{ijk}b^j{}_\b\left(p\phi^{k\b}+q\Phi^{k\b}\right)
  +(u\phi^{j\b}-v\Phi^{j\b})\frac{1}{2}\Th_{ji\b} \, ,     \nn\\
&&\hcK_i=\cK_i-\nabla_\b\Phi_i{}^\b
  -\ve_{ijk}b^j{}_\b\phi^{k\b} \, ,                        \nn\\
&&\hat D^\a=D^\a+b^n{}_0\pi_n{^\a}+\om^n{}_0\Pi_n{}^\a\, . \nn
\eea
The consistency conditions of the secondary constraints  are
identically satisfied, and the Hamiltonian consistency procedure is
thereby completed.

Regarding the classification of constraints, we note that
$(\p_i{^0},\Pi_i{^0},\pi^0)$ and $(\hcH_i,\hcK_i,\pd_\a\pi^\a)$ are
first class, while $(\phi_i{^\a},\Phi_i{^\a})$ are second class.

Using the well-established Castellani procedure \cite{15},  we obtain
the canonical generator:
\bea
&&G=-G_1-G_2-G_3 \, ,                                      \nn\\
&&G_1:=\dot\xi^\r\left(b^i{}_\r\pi_i{}^0
       +\om^i{}_\r\Pi_i{}^0+\pi^0A_\r\right)               \nn\\
&&\qquad\,
  +\xi^\r\left[b^i{}_\r\hat\cH_i+\om^i{}_\r\hat\cK_i
  +(\pd_\r b^i_0)\pi_i{}^0
  +(\pd_\r\om^i{}_0)\Pi^i{}_0+(\pd_\r A_0)\pi^0 \right]\, ,\nn\\
&&G_2:=\dot{\th^i}\Pi_i{}^0+\th^i\left[\hcK_i
  -\ve_{ijk}\left(b^j{}_0\pi^{k0}
  +\om^j{}_0\Pi^{k0}\right)\right]\, ,                     \nn\\
&&G_3:=\dot{\l}\pi^0-\l\pd_\a\pi^\a \, ,                   \nn
\eea
where the integration symbol $\int d^2 x$ is omitted for simplicity.
The action of $G$ on the fields is defined by the Poisson bracket
operation, $\bar\d\phi:=\{\phi,G\}$. As one can verify, $\bar\d\phi$
coincides with the combination of Poincar\'e plus $U(1)$ gauge
transformations on shell.


\begin{thebibliography}{99} 

\bibitem{1} For a review of the subject and an extensive list of
  references, see: S. Carlip, {\it Quantum Gravity in 2+1 Dimensions\/}
  (Cambridge University Press, Cambridge, England, 1998);
  Conformal Field Theory, (2+1)-dimensional Gravity, and the BTZ Black
  Hole, Class. Quant. Grav. {\bf 22} (2005) R85-R124. \vsm
\bibitem{2} M. Ba\~nados, C. Teitelboim and J. Zanelli, The Black Hole
  in Three-Dimensional Spacetime, Phys. Rev. Lett. {\bf 69} (1992) 1849;\\
  M. Ba\~nados, M. Henneaux, C. Teitelboim and J. Zanelli, Geometry of
  2+1 Black Hole, Phys. Rev. {\bf D 48} (1993) 1506. \vsm
\bibitem{3} E. W. Mielke, P. Baekler, Topological gauge model of
  gravity with torsion, Phys. Lett. {\bf A 156} (1991) 399;\\
  P. Baekler, E. W. Mielke, F. W. Hehl,
  Dynamical symmetries in topological 3D gravity with torsion,
  Nuovo Cim. {\bf B 107} (1992) 91. \vsm
\bibitem{4} F. W. Hehl, Four lectures on Poincar\'e gauge theory,
  Proceedings of the 6th Course of the School of Cosmology and
  Gravitation, on Spin, Torsion, Rotation and Supergravity, held in
  Ericce, Italy, 1979, eds. P. G. Bergmann, V. de Sabata (Plenum, New
  York, 1980) p. 5;\\ E. W. Mielke, {\it Geometrodynamics of Gauge
  Fields\/} -- On the geometry of Yang-Mills and gravitational gauge
  theories (Akademie-Verlag, Berlin, 1987). \vsm
\bibitem{5}  M. Blagojevi\'c, {\it Gravitation and gauge symmetries\/}
  (IoP Publishing, Bristol, 2002); Three lectures on Poincar\'e gauge
  theory, SFIN {\bf A 1} (2003) 147 [gr-qc/0302040].\vsm
\bibitem{6} A. Garc\'\i a, F. W. Hehl, C. Heinecke and A. Mac\'\i as,
  Exact vacuum solutions of (1+2)-dimensional Poincar\'e gauge theory:
  BTZ solution with torsion, Phys. Rev. {\bf D 67} (2003) 124016;\\
  M. Blagojevi\'c  and M. Vasili\'c,
  3D gravity with torsion as a Chern-Simons gauge theory,
  Phys. Rev. {\bf D 68} (2003) 104023;\\
  E. W. Mielke and A. A. R. Maggiolo, Rotating black hole
  solution in a generalized topological 3D gravity with torsion,
  Phys. Rev. {\bf D 68} (2003) 104026.\vsm
\bibitem{7} M. Blagojevi\'c and B. Cvetkovi\'c, Canonical structure of
  3D gravity with torsion, in: {\it Progress in General Relativity and
  Quantum Cosmology\/}, vol. 2, ed. Ch. Benton (Nova Science Publishers,
  New York, 2006), pp. 103-123 [gr-qc/0412134];
  Black hole entropy in 3D gravity with torsion, Class. Quantum Grav.
  {\bf 23} (2006) 4781. \vsm
\bibitem{8} Y. Obukhov, New solutions in 3D gravity,
  Phys. Rev. {\bf D 68} (2003) 124015;\\
  S. Cacciatori, M. Caldarelli, A. Giacomini, D. Klemm and
  D. Mansi, Chern-Simons formulation of three-dimensional gravity with
  torsion and nonmetricity, J. Geom. Phys. {\bf 56\/} (2006) 2623;\\
  D. Klemm, G. Tagliabue, The CFT dual of AdS gravity with torsion,
  Class. Quant. Grav. {\bf 25} (2008) 035011. \vsm
\bibitem{9}  A. Giacomini, R. Troncoso and S. Willison,
  Three-dimensional supergravity reloaded,
  Class. Quantum Grav. {\bf 24} (2007) 2845;\\
  B. Cvetkovi\'c and M. Blagojevi\'c,
  Supersymmetric 3D gravity with torsion: asymptotic symmetries,
  Class. Quantum Grav. {\bf 24} (2007) 3933; Stability of 3D black hole
  with torsion, Mod. Phys. Lett. {\bf A 22} (2007) 3047.\vsm
\bibitem{10} G. Clement, Classical solutions in three-dimensional
  Einstein-Maxwell cosmological gravity, Class. Quant. Grav. {\bf 10}
  (1993) L49-L54;\\
  J. F. Chan, K. K. Chan and R. B. Mann, Interior Structure of a
  Charged Spinning Black Hole in (2 + 1)-Dimensions, Phys. Rev.
  {\bf D 54} (1996) 1535;\\
  M. Cataldo and P. Salgado, Static Einstein-Maxwell Solutions in
  2+1 dimensions, Phys. Rev. {\bf D 54} (1996) 2971;\\
  D. H. Park and S. Yang, Geodesic Motions in 2+1 Dimensional
  Charged Black Holes, Gen. Rel. Grav. {\bf 31} (1999) 1343;\\
  C. Martinez, C. Teitelboim and J. Zanelli, Charged Rotating
  Black Hole in Three Spacetime Dimensions, Phys. Rev. {\bf D 61}
  (2000) 104013. \vsm
\bibitem{11}  E. W. Hirschmann and D. L. Welch, Magnetic solutions
  to 2+1 gravity,  Phys. Rev. {\bf D 53} (1996) 5579;\\
  Y. Kiem and D. Park, Magnetically charged solutions via an analog
  of the electric-magnetic duality in (2+1)-dimensional gravity theories,
  Phys. Rev. {\bf D 55} (1997) 6112;\\
  M.  Cataldo, J. Cris\'ostomo, S. del Campo and P. Salgado, On
  Magnetic solution to 2+1 Einstein-Maxwell gravity, hep-th/0401189;\\
  R. Olea, Charged rotating black hole formation from thin shell
  collapse in three dimensions,
  Mod. Phys. Lett. {\bf A 20} (2005) 2649.\vsm
\bibitem{12} M. Kamata and T. Koikawa, The electrically charged
  BTZ black hole with self (anti-self) dual Maxwell field,
  Phys. Lett. {\bf B 353} (1995) 196;\\
  G. Cl\'ement, Spinning charged BTZ black holes and self-dual
  particle-like solutions, Phys. Lett. {\bf B 367} (1996) 70;\\
  K. Chan, Comment on the calculation of the angular momentum and
  mass for the (anti-)self dual charged spinning BTZ black hole,
  Phys. Lett. {\bf B 373} (1996) 296;\\
  M. Kamata and T. Koikawa, (2+1) dimensional charged black
  hole with (anti-)self dual Maxwell fields,
  Phys. Lett. {\bf B 391} (1997) 87;\\
  M. Cataldo and P. Salgado, Three dimensional extreme black
  hole with self (anti-self) dual Maxwell field,
  Phys. Lett. {\bf B 448} (1999) 20. \vsm
\bibitem{13} G. Cl\'ement, Classical solutions of gravitating
  Chern-Simons electrodynamics, gr-qc/9406052;\\
  S. Fernando and F. Mansouri, Rotating Charged Solutions
  to Einstein-Maxwell-ChernSimons Theory in 2+1 Dimensions,
  Commun. Math. Theor. Phys. {\bf 1} (1998) 14;\\
  T. Dereli and Yu. N. Obukhov, General analysis of self-dual
  solutions for the Einstein-Maxwell-Chern-Simons theory in (1+2)
  dimensions, Phys. Rev. {\bf D 62} (2000) 024013.\vsm
\bibitem{14} M. Cataldo, Azimuthal electric field in a static
  rotationally symmetric (2+1)-dimensional spacetime,
  Phys. Lett. {\bf B 52} (2002) 143.\vsm
\bibitem{15} L. Castellani, Symmetries of constrained Hamiltonian
  systems, Ann. Phys. (N.Y) {\bf 143} (1982) 357.\vsm
\bibitem{16} M. Blagojevi\'c and B. Cvetkovi\'c, Self-dual Maxwell
  field in 3D gravity with torsion, Phys. Rev. {\bf D 78}, 044037 (2008) [arXiv:0805.3627]

\end{thebibliography}
\end{document}